\documentstyle[12pt,aps,prl,psfig]{revtex}
\begin{document}
\draft 
\tighten

\title{Quantitative tests of mode-coupling theory for fragile and
strong glass-formers}

\author{Walter Kob$^{(1)}$, Markus Nauroth$^{(2)}$, and Francesco
Sciortino$^{(3)}$}

\address{
$^{(1)}$ Laboratoire des Verres, Universit\'e Montpellier II, F-34095
Montpellier, France\\
$^{(2)}$ Institut f\"ur Physik, Johannes--Gutenberg--Universit\"at,
Staudinger Weg 7, D--55099 Mainz, Germany\\
$^{(3)}$ Dipartimento di Fisica and Istituto Nazionale
per la Fisica della Materia, Universit\'a di Roma {\it La Sapienza},
P.le Aldo Moro 2, I-00185 Roma, Italy
}

\date{17. September, 2001}

\maketitle

\begin{abstract}
We calculate for a binary mixture of Lennard-Jones particles the time
dependence of the solution of the mode-coupling equations in which the
full wave vector dependence is taken into account. In addition we also
take into account the short time dynamics, which we model with a simple
memory kernel. We find that the so obtained solution agrees very well
with the time and wave vector dependence of the coherent and incoherent
intermediate scattering functions as determined from molecular dynamics
computer simulations. Furthermore we calculate the wave vector dependence
of the Debye-Waller factor for a realistic model of silica and compare
these results with the ones obtained from a simulation of this model. We
find that if the contribution of the three point correlation function to
the vertices of the memory kernel are taken into account, the agreement
between theory and simulation is very good. Hence we conclude that mode
coupling theory is able to give a correct quantitative description of
the caging phenomena in fragile as well as strong glass-forming liquids.

\end{abstract}


\section{Introduction}
\label{sec1}

In the last decade our understanding of the dynamics of supercooled
liquids has made significant progress~\cite{proceedings}. In particular
it has been shown that for fragile glass-formers the bend one observes
if one plots the logarithm of the viscosity as a function of the
inverse temperature can be explained very well by means of the so-called
mode-coupling theory of the glass transition (MCT)~\cite{gotze99}. In the
vicinity of this bend the dynamics of the system changes qualitatively
in that the particles start to experience strong caging effects,
i.e. they are temporarily trapped by the particles that surround
them. MCT gives a self-consistent description of the dynamics of the
particles inside this cage as well as how the particles leave this cage,
i.e. of the structural relaxation of the supercooled liquid. In the past
the predictions of this theory have been checked in many experiments
as well as computer simulations and it was found that MCT is indeed
able to give a qualitatively correct description of the relaxation
dynamics~\cite{gotze99}.

However, in principle the theory is supposed to give not only a
qualitatively description of the relaxation dynamics of supercooled
liquids, but also a quantitative one, {\it if} the static properties
of the system are known with sufficiently high precision. This
attractive feature originates directly from the way the theory is
(or can be) derived, namely the Mori-Zwanzig formalism in which one
obtains equations of motion for slow variables which involve their
{\it static} values. Thus, once these static values are known one
can, in principle, determine their time dependence. In particular it
is possible to calculate from the knowledge of the static structure
factor the time dependence of the coherent and incoherent intermediate
scattering functions, $F(q,t)$ and $F_s(q,t)$, respectively, where
$q$ is the wave vector. Unfortunately, in the past these type of
calculations have been done only for very few systems, since one
the one hand they are quite involved and on the other hand they
require as input structural data with very high quality (better than
1\%)~\cite{barrat90,gotze91,fuchs92a,fuchs92b,nauroth97,water,winkler00}.
In the present paper we expand these type of calculations in two
directions. On the one hand we solve for a simple glass-forming system,
a binary Lennard-Jones mixture (BMLJ), the full time and wave vector
dependence of the MCT equations, including a realistic short time
dynamics, and compare them with results from computer simulations of
the same system. On the other hand we calculate the $q-$dependence of
the nonergodicity parameters (NEP) for silica (SiO$_2$), a glass-former
whose structure is given by an open tetrahedral network and who is the
prototype of a strong glass-former, and compare also these results with
the ones from simulations of the same system.

\section{Theory}
\label{sec2}

In this section we summarize the MCT equations that are needed to
calculate the quantities discussed in section~\ref{sec4}. In order to
keep this presentation as simple as possible we will discuss only the
equation for the case of a one-component system, although in reality we
have used the equations for a two-component system, since the BMLJ as
well as SiO$_2$ belong to this class. The full binary equations can be
found in Refs.~\cite{barrat90,nauroth_phd,sciortino01}.

The intermediate scattering function can be defined by
$F(q,t)=\langle \delta \rho({\bf q},0) \delta \rho({\bf q},t)\rangle$
where $\delta \rho$ are the density fluctuations. $F(q,t)$ obeys the exact
equation of motion

\begin{equation}
\ddot{F}(q,t)+\Omega^2(q)F(q,t)+\int_0^tM(q,\tau)\dot{F}(q,t-\tau)d\tau = 0
\label{eq1}
\end{equation}
where the frequency $\Omega^2$ is given by $\Omega^2=q^2 k_B T/(m
S(q))$. Here $m$ and $S(q)$ are the mass of the particles and the static
structure factor respectively. The function $M(q,t)$ in Eq.~(\ref{eq1})
is the so-called memory function and it is useful to write it as follows:

\begin{equation}
M(q,t) =  M^{REG}(q,t)+
\left\{M^{MCT}[F(k,t)](q)-M^{MCT}[F^B(k,t)](q)\right\}.
\label{eq2}
\end{equation}

Here $M^{REG}(q,t)$ is that part of the memory function which is
responsible for the dynamics of the system at very short times,
i.e. after the particles have left the ballistic regime. The functional
$M^{MCT}[F(k,t)]$ is the usual memory kernel of MCT which depends on
the static structure factor as well as on the three point correlation
function $c_3({\bf q}, {\bf k})$~\cite{gotze99}. Also it contains
a short time part. But since we want to describe this time regime
by means of $M^{REG}(q,t)$, we have to subtract out this part from
$M^{MCT}[F(k,t)]$. This is done in the last term of Eq.~(\ref{eq2}),
where $F^B$ is a function which decays rapidly to zero, but has the
correct behavior at short times~\cite{nauroth_phd}.

For the regular memory function $M^{REG}(q,t)$ one can make different
type of Ansatzes. One which seems to work well at high temperatures is
given by~\cite{tankeshwar}

\begin{equation}
M^{REG}(q,t)=\alpha(q) /\cosh(\beta(q) t).
\label{eq3}
\end{equation}
Here $\alpha(q)$ and $\beta(q)$ are parameters which can be
calculated via sum rules from the static structure factor
and other static quantities which can be measured in a computer
simulation~\cite{nauroth_phd,boon80}. Hence they are {\it not}
adjustable fit parameters.

Eqs.~(\ref{eq1})-(\ref{eq3}) are a self-contained set of equations of
motion from which one thus can calculate the full time and wave vector
dependence of $F(q,t)$, and similar equations exist for the incoherent
intermediate scattering function $F_s(q,t)$. We have solved these
equations by discretizing $q$-space into 100 points that covered the
$q-$range up to 3-4 times the location of the main peak in $S(q)$.

\section{Models and details of the simulations}
\label{sec3}
The first model investigated is a 80:20 mixture of Lennard-Jones
particles. In the following we will call the majority and minority
species A and B particles, respectively. Both of them have the same
mass $m$ and they interact via a potential $V_{\alpha\beta}=
4\epsilon_{\alpha\beta}[(\sigma_{\alpha\beta}/r)^{12}-
(\sigma_{\alpha\beta}/r)^6]$, $\alpha,\beta\in \{\rm A,B\}$. The
parameters $\epsilon_{\alpha\beta}$ and $\sigma_{\alpha\beta}$
are given by $\epsilon_{\rm AA}=1.0$, $\sigma_{\rm AA}=1.0$,
$\epsilon_{\rm AB}=1.5$, $\sigma_{\rm AB}=0.8$, $\epsilon_{\rm BB}=0.5$, and
$\sigma_{\rm BB}=0.88$. This potential is truncated and shifted at a distance
$\sigma_{\alpha\beta}$. In the following we will use $\sigma_{\rm AA}$
and $\epsilon_{\rm AA}$ as the unit of length and energy, respectively
(setting the Boltzmann constant $k_{\rm B}=1.0$). Time will be measured
in units of $\sqrt{m\sigma_{\rm AA}^2/48\epsilon_{\rm AA}}$. In the
past the structural and dynamical properties of this system have been
studied in great detail~\cite{kob_lj,gleim98}. More detail on this can
be found in Ref.~\cite{kob_99}. 

For the present work we only needed to determine the three point
correlation function $c_3$ since the time and temperature dependence
of $F(q,t)$ and $F_s(q,t)$, as well as the one of $S(q)$, can be
found in the mentioned literature. For this we simulated a system of
800 A particles and 200 B particles in a box with volume $(9.4)^3$.
The total time of this simulation was about $10^8$ time steps from
which we obtained roughly 12,000 independent configurations. This large
number was necessary to determine $c_3$ with sufficient precision.
Due to this large computational effort we did this calculation only for
one temperature, $T=1.0$. Thus in the following we will assume that the
temperature dependence of $c_3$ is weak.

The second model we study is amorphous silica, SiO$_2$. For this we use
the potential proposed by van Beest {\it et al.} which has the functional
form~\cite{beest90}

\begin{equation}
\phi_{\alpha \beta}(r)=
\frac{q_{\alpha} q_{\beta} e^2}{r} +
A_{\alpha \beta} \exp\left(-B_{\alpha \beta}r\right) -
\frac{C_{\alpha \beta}}{r^6}\quad \alpha, \beta \in
[{\rm Si}, {\rm O}].
\label{eq4}
\end{equation}

The values of the constants  $q_{\alpha}, q_{\beta}, A_{\alpha
\beta}, B_{\alpha \beta}$, and $C_{\alpha \beta}$ can be found in
Ref.~\cite{beest90}. The potential has been truncated and shifted at
5.5~\AA. In the past it has been shown that this potential is able to
give a reliable description of silica in its molten phase as well as in
the glass (see~\cite{horbach99,horbach01,saika01} and references therein). For
the present calculations to determine $c_3$ we used 600 ions in a box with
volume (20.4\AA)$^3$. The total length of the simulation was $2\times
10^7$ time steps, from which we obtained at 4000~K around 2000 independent
configurations.

\section{Results}
\label{sec4}
\begin{figure}
\psfig{figure=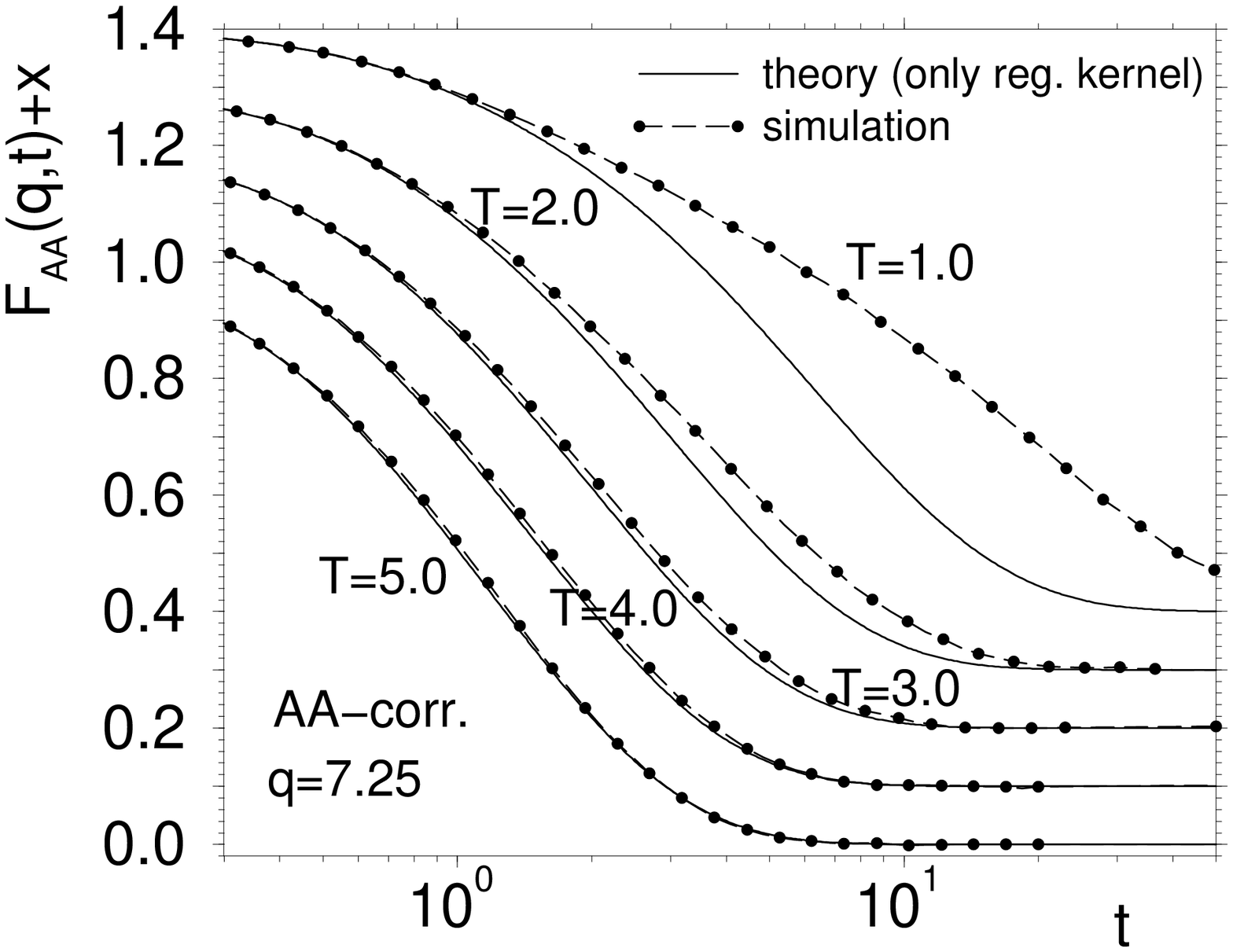,width=10.0cm,height=8.5cm}
\caption{Time dependence of the coherent intermediate scattering function for the
A particles in the BMLJ system 
for different temperatures as obtained from the simulation (dashed
lines with symbols) and as predicted from the theory if the MCT kernel is not
taken into account (solid lines). For clarity some of the curves have been 
shifted vertically.}
\label{fig1}
\end{figure}

We start by considering first the dynamics of the BMLJ at intermediate
and high temperatures. In this $T-$range it can be expected that the
effect of the memory kernel of MCT is not relevant and thus we will
set it to zero. In Fig.~\ref{fig1} we show the time dependence of
$F_{\rm AA}(q,t)$ for various temperatures. The wave vector is 7.25,
the location of the main peak in $S_{\rm AA}(q)$. The dashed lines with
symbols are the result of the simulation whereas the full lines are the
prediction of the theory. As can be seen, the theory works very well at
high temperatures but starts to break down at intermediate temperatures
in that it underestimated the correlation function at intermediate
times. Thus we see that even at the intermediate temperature $T=1.0$,
which is more than twice the MCT temperature $T_c=0.435$, cage effects
become important.

In order to see whether the memory kernel $M^{MCT}$ is able to take
into account these effects we have solved Eqs.~(\ref{eq1})-(\ref{eq3})
by taking now into account also this contribution to the memory function
$M(q,t)$. In doing this we had to face a problem which we had encountered
already some time ago~\cite{nauroth97}, namely that MCT is not able
to predict reliably the value of the critical temperature $T_c$. For
the BMLJ the simulations show that $T_c\approx 0.435$~\cite{kob_lj},
whereas the theory predicts a value around 0.92~\cite{nauroth97}. This
means that the theory is not able to predict correctly the {\it absolute}
value of the time scale for the $\alpha-$relaxation, although it is able
to predict the shape of the correlation functions (see below). Therefore
we had to use {\it one} adjustable parameter, a temperature which we
will denote by $T_f$, which is the temperature at which the vertices
in the MCT-functional $M^{MCT}$ are evaluated. The value of $T_f$ was
adjusted such that the time scale for $F_{\rm AA}(q,t)$ for $q=7.25$
was reproduced correctly.

\begin{figure}
\psfig{figure=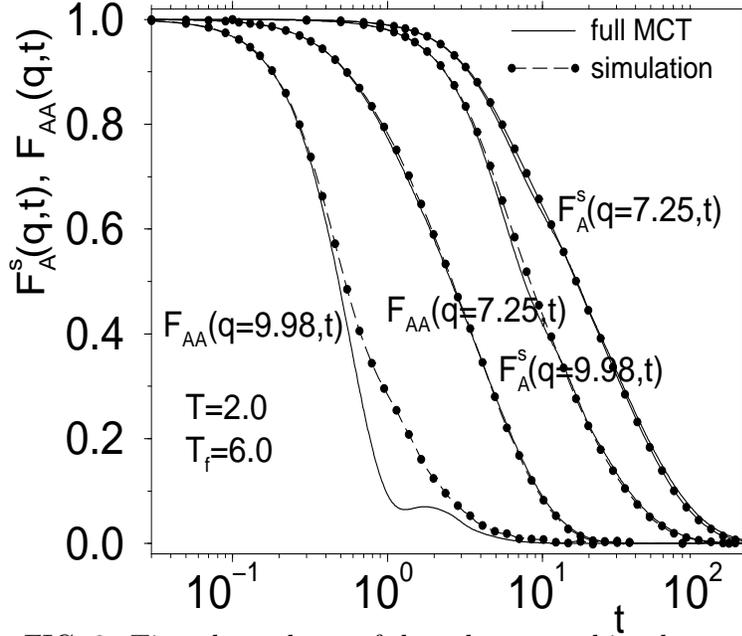,width=10.0cm,height=8.5cm}
\caption{Time dependence of the coherent and incoherent intermediate
scattering function of the BMLJ system for two wave vectors at $T=2.0$. The dashed line
with the symbols are the results from the simulation and the solid lines
are the prediction of the theory in which the MCT kernel has been taken
into account.}
\label{fig2}
\end{figure}

In Fig.~\ref{fig2} we show the time dependence of the coherent as well as
the incoherent intermediate scattering function for $q=7.25$ and $q=9.98$,
the location of the first peak and the first minimum in $S_{\rm AA}(q)$,
respectively. The temperature is 2.0, i.e. a value for which we find that
the {\it regular} memory kernel is no longer able to give a good description of
the relaxation dynamics (see Fig.~\ref{fig1}). From this figure we see
that in general the agreement between the simulation and the theory is
very good in that the shape of the curves as well as their position is
correctly predicted. (The discrepancy found for $F_{\rm AA}(q=9.98,t)$,
where the theory predicts a pronounced shoulder at around $t=2$
whereas the simulation shows only a weak shoulder in that time regime,
is probably related to the fact that the description of the short time
dynamics is not yet completely adequate~\cite{nauroth_phd}). Thus we
conclude that MCT is indeed able to give a correct description of the
relaxation dynamic of the system at intermediate temperatures, i.e. at
temperatures where the cage effect starts to become noticeable.

\begin{figure}
\psfig{figure=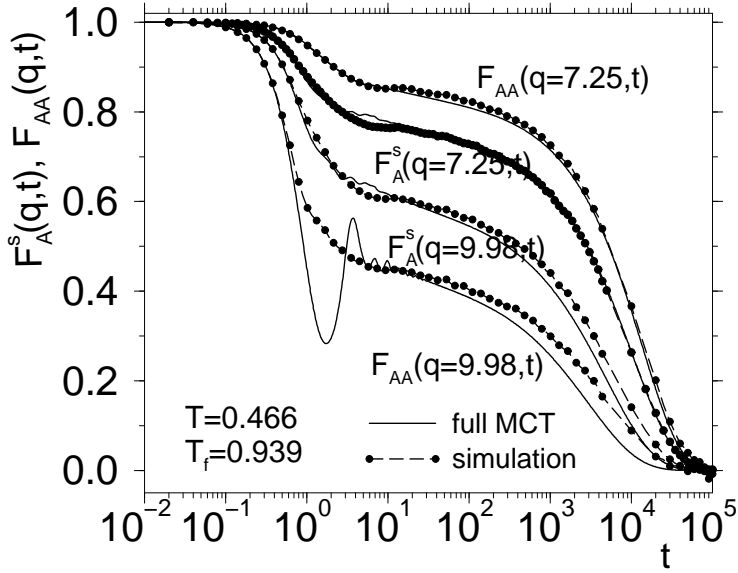,width=10.0cm,height=7.5cm}
\caption{Same as figure 3 but now for $T=0.466$.}
\label{fig3}
\end{figure}

We now check whether this conclusion is also correct if the temperature
is so low that the cage effect becomes very important. For this we have
solved the MCT equations for $T=0.466$, i.e. a temperature for which the
relaxation dynamics is about $10^4$ times slower than the one at high
$T$. The time dependence of $F_{\rm AA}(q,t)$ and of $F_{\rm A}^s(q,t)$
as predicted from the theory is shown in Fig.~\ref{fig3}. Also included
are the results from the computer simulations from Ref.~\cite{kob_lj}. As
in the case of intermediate temperatures, Fig.~\ref{fig2}, we find
that also for this $T$ the agreement between theory and simulation is
very good. The main discrepancy is again seen for $F_{\rm AA}(q,t)$ at
$q=9.98$, and the reason for it is the same as the one given above. All
in all we thus conclude that for this system the theory is indeed able
to predict the full time and wave vector dependence of the coherent and
incoherent scattering function.

The temperature dependence of the relaxation time of the BMLJ system
shows significant deviations from an Arrhenius law~\cite{kob_lj}. As
mentioned above, these deviations are believed to be related to a change
in the transport mechanism of the particles which show a hopping type
of motion at low temperatures whereas at high $T$ they show a more
collective/flow-like behavior. It is of interest that recently it
has been suggested that even silica shows such a crossover in the transport
mechanism, although this crossover occurs at relatively high temperatures
(around 3300K)~\cite{horbach99}. Therefore one might ask whether MCT
is able to give a reliable description of the relaxation dynamics of
this important glass-forming system also. Note that from a structural
point of view the BMLJ system and amorphous silica are very different,
since the former one resembles to the random close packing of hard
spheres whereas the latter is given by an open network structure
similar to the continuous random tetrahedral network proposed long
time ago by Zachariasen~\cite{zachariasen32}. Since, as pointed out
in the Introduction, MCT uses only structural information to predict
the dynamics, it is of great interest to see whether the theory is also
able to give a correct quantitative description if the structure is very
different from the one of closed packed hard spheres.

To check this we have calculated for amorphous silica the wave vector
dependence of the nonergodicity parameters, i.e. the height of the plateau
in the intermediate scattering function at intermediate times (see, e.g.,
Fig.~\ref{fig3}). Before we discuss the results, we have to mention a
technical point which makes the calculation of the NEP for the case of
silica much harder than for the case of the BMLJ. In Sec.~\ref{sec2}
we mentioned that the memory kernel of the MCT contains only static
quantities, namely the static structure factor $S(q)$ and the three point
correlation function $c_3({\bf q}, {\bf k})$. From a simulation it is
quite easy to determine $S(q)$ with high precision. For the function
$c_3$ this is, however, not the case, since due to the two vectorial
arguments the statistics for this quantity is very bad. Therefore we had
to make very long simulations in order to determine $c_3$ with sufficient
accuracy. More details on this can be found in Ref.~\cite{sciortino01}.

In the following we will discuss the results for the NEP for the BMLJ as
well as for the case of silica. Since all of the results presented in
Figs.~\ref{fig1}-\ref{fig3} were obtained with the approximation that
$c_3\equiv 0$, one has of course to check whether or not they do not
change if this assumption is not made. We mention, however, already here,
that some time ago Barrat {\it et al.} showed that this approximation is
very good for the case of a soft-sphere system, i.e. a system which is
relatively similar to the BMLJ considered here~\cite{barrat89}. Whether
this result holds also for the case of a system with an open network
structure has, however, so far not been investigated.  We also mention
that in order to calculate the NEPs it is not necessary to introduce any
fit parameter of any kind. The only input to the data are the partial
structure factors~\cite{gotze99}. Thus for this type of calculation the
above discussed problem with the $T_f$ does not exist.

\begin{figure}
\psfig{figure=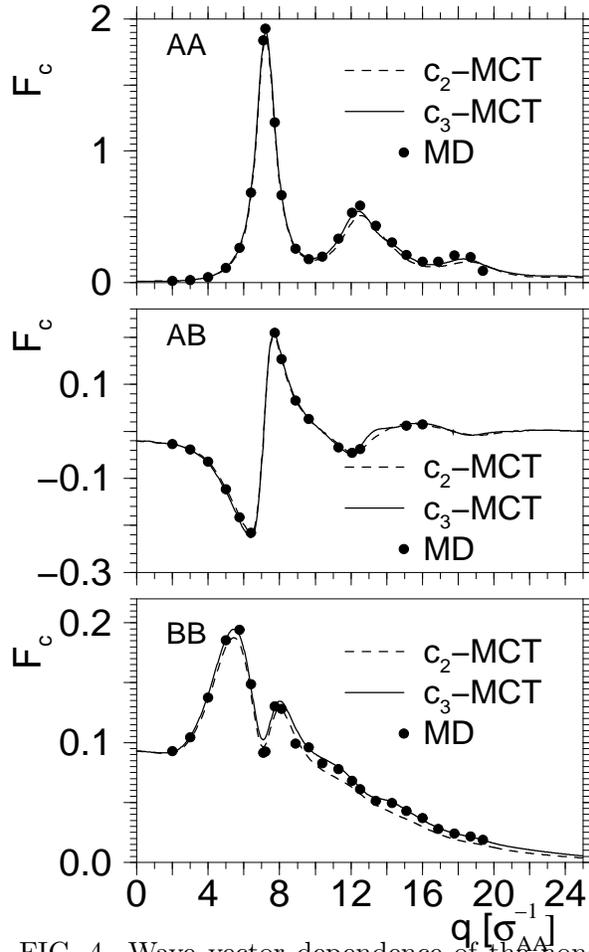,width=8.0cm,height=12.5cm}
\caption{Wave vector dependence of the nonergodicity parameters for the BMLJ
system. The circles are the result of the simulation, the dashed line is the
theoretical prediction if the three point correlation function is set to zero, and
the full line is the theoretical prediction if this function is taken into
account.}
\label{fig4}
\end{figure}

In Fig.~\ref{fig4} we show the wave vector dependence of the NEP for the
coherent functions. (Note that since this is a binary system, there are
three of them. Furthermore we mention that for reasons of convenience
we show the NEP multiplied by the corresponding partial structure
factors.) In each panel we show three curves: The circles are the result
from the simulation published in Ref.~\cite{gleim98}. The
dashed and full line is the theoretical result for the cases that $c_3$
is set equal to zero and $c_3\neq 0 $, respectively. First of all this
figure shows that the theory is able to reproduce with excellent accuracy
the data of the simulation without any adjustable parameter. Furthermore
we recognize from Fig.~\ref{fig4} also that the theoretical prediction
hardly depends on whether or not $c_3$ is taken into account, in agreement
with the finding of Barrat {\it et al.}~\cite{barrat89}.

\begin{figure}
\psfig{figure=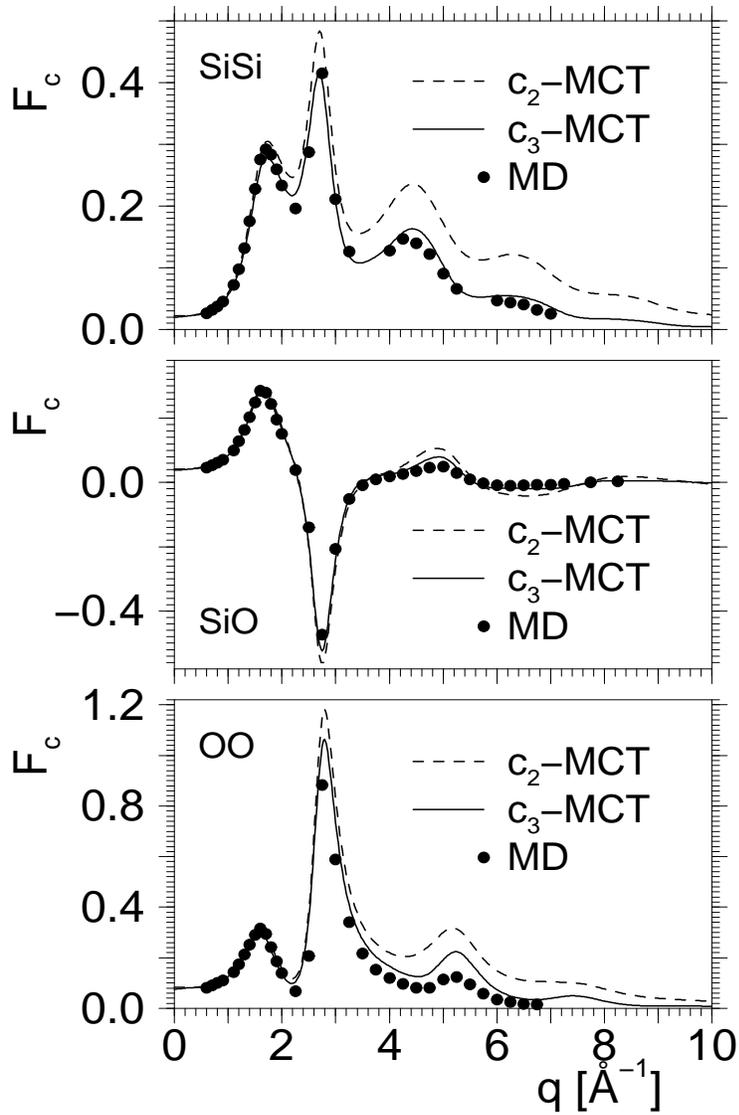,width=10.0cm,height=14.5cm}
\caption{Same as Fig.~\protect\ref{fig4} but for the case of silica.}
\label{fig5}
\end{figure}

For silica the situation is quite different as can be inferred from
Fig.~\ref{fig5} where we show the wave vector dependence of the NEP
for this system. We see that in this case the theoretical prediction
for $c_3=0$ differs strongly from the one if this function is taken
into account. Thus we find that for the case of a network structure
the contribution of the three point correlation function to the memory
function is very important. It is remarkable that if the contributions
of $c_3$ are taken into account, the theoretical prediction agrees
very well with the result of the simulation, which were presented in
Ref.~\cite{horbach01}. Thus we conclude that the theory is also able to
give a quantitative correct prediction for this type of glass-former.

\section{Summary}
\label{sec5}

The goal of this work was to check to what extend the mode-coupling theory
of the glass transition is able to give a correct {\it quantitative}
description of the dynamics of glass-forming liquids. This was done for
two very different types of systems: A binary mixture of Lennard-Jones
particles, whose structure is similar to the one of a close packing of
hard spheres and whose temperature dependence of relaxation times makes
it a glass-forming liquid of intermediate fragility. On the other hand
we have studied silica, who has a open network structure and which, in
the temperature region where experiments are feasible, is considered
to be the prototype of a strong glass-former.

For the BMLJ system we have solved the MCT equations to obtain the
full time and wave vector dependence of the relaxation dynamics. In
particular we have also included a realistic Ansatz for the dynamics at
short times so that the theoretical curves should be able to give also
a good description in this time regime. By comparing the theoretical
curves for $F(q,t)$ and $F_s(q,t)$ with the ones obtained from computer
simulations of the same system, we find that cage effects become
noticeable already at relatively high temperatures. The theory is able to
give a very reliable quantitative description of the relaxation dynamics
for all temperatures considered. The only discrepancy found is probably
related to the fact that our understanding of the dynamics at {\it short}
times is still incomplete.

For the case of silica we have calculated the $q-$dependence of the
nonergodicity parameters. We have found that for this system it is
important to include in the evaluation of the memory function also those
contributions that stem from three point correlation functions. Most
probably this finding is related to the open network structure of
the system. We find that once $c_3$ is taken into account the prediction of the 
theory for the NEP are in very good agreement with the results from computer
simulations.

In summary we have shown that MCT is able to give a very good {\it
quantitative} description of the relaxation dynamics of fragile as well
as strong liquids, at least at high and intermediate temperatures. Thus
we conclude that this theory is able to rationalize {\it at least} the
first few decades of the slowing down of a very large class of 
glass-forming liquids on a quantitative level.

Acknowledgments: We thank L. Fabbian for contributing to the early
development of this research, and W. G\"otze for useful discussions
on this work. Part of this work was supported by the DFG through
SFB~262. F.S. acknowledges support from INFM-PRA-HOP and MURST-PRIN2000.
W.~K. thanks the Universit\'a La Sapienza for a visiting professorship
during which part of this work was carried out.


\section*{Figures}

\vspace*{-7mm}

\end{document}